\documentclass[10pt,twoside]{Lowx2011}
\usepackage{epsf,amsmath}
\usepackage{graphicx}

\begin{document}

\title{Results from the Pierre Auger Observatory}

\author{Ruben~Concei\c{c}\~ao \\ \\
{\it for the Pierre Auger Collaboration} \\ \\
LIP, Av. Elias Garcia, 14-1, 1000-149 Lisbon, Portugal\\
Observatorio Pierre Auger, Av. San Mart\'in Norte 304,\\
 (5613) Malarg\"{u}e, Mendoza, Argentina \\
(Full author list: http://www.auger.org/archive/author\_2011\_06.html) \\
E-mail: ruben@lip.pt}

\maketitle

\begin{abstract}
\noindent The Pierre Auger Observatory is currently the largest observatory of Ultra High Energy Cosmic Rays. Having more data collected than any previous experiment and using a hybrid technique, it can provide important information to unveil the origin and composition of Ultra High Energy Cosmic Rays.
Here, we report some results of the Pierre Auger Observatory, namely on the energy spectrum, average depth of the shower maximum and its fluctuations (both sensitive to primary mass composition) and number of muons at ground.
\end{abstract}



\markboth{\large \sl \hspace*{0.25cm}R.~Concei\c{c}\~ao for the Pierre Auger Collaboration 
\hspace*{0.25cm} Low-$x$ Meeting 2011} {\large \sl \hspace*{0.25cm} Results from the Pierre Auger Observatory}

\section{Introduction} 
The Pierre Auger Observatory was built to unveil the origin and composition of the most energetic particles known, the Ultra High Energy Cosmic Rays (UHECRs). These particles are known to arrive at Earth with a very scarce flux. Fortunately they interact with the atmosphere producing huge particle showers that can be detected either at the ground or by observing the radiation emitted by the shower as it travels through the atmosphere.
The Pierre Auger Observatory \cite{PAO} is currently the largest UHECR observatory on Earth, covering $3000$ km$^{2}$ of the high plateau of Pampa Amarilla, near Malarg\"ue, in Argentina. It is composed by about $1660$ water Cherenkov stations, spaced by $1.5$ km and $24$ fluorescence telescopes that overlook the atmosphere above the array \cite{FD}. The telescopes, which are based on the modified Schmidt optics, have a field of view of $30^\circ$ in azimuth and of $28.6^\circ$ in elevation. 

The Observatory is taking data since $2004$ and the baseline configuration was achieved in mid $2008$. The Observatory applies two main techniques to observe Extensive Air Showers (EAS): sampling of charged particles that arrive at the ground level and following the longitudinal development of the shower by detecting the fluorescence light emitted by nitrogen molecules excited by the shower particles traveling through the atmosphere. While the Surface Detector (SD) has a duty cycle of almost $100$\% the Fluorescence Detector (FD) can only operate at moonless clear nights, thus having a duty cycle of $\sim 12$\%. During this time the Observatory can operate in a hybrid mode allowing us to extract more information about each individual shower. The hybrid technique offers several advantages, for instance it allows us to considerably improve the geometric shower reconstruction. But perhaps the most important feature is that it allows us to calibrate the signal measured with the SD to the energy measured by the FD, which is almost independent of hadronic interaction models.

The energy measured by the SD is obtained through the measurement of the signal at the ground at $1000$ meters from the shower core, $S(1000)$. This distance was chosen as it was proven in simulation to be the less sensitive to shower-to-shower fluctuations, primary composition and high energy hadronic interaction models. However the signal at the ground depends naturally on the amount of matter traversed by the shower and consequently on the shower zenith angle, $\theta$. Therefore $S(1000)$ is converted to a reference angle $(\theta = 38^\circ)$, $S_{38^\circ}$. This conversion can be obtained from the SD data itself through the Constant-Intensity-Cut method \cite{CIC}. On the other hand, the integral of the fluorescence profile measured by the FD gives a \emph{quasi}-calorimetric measurement of the primary particle energy, which only has to be corrected for the \emph{invisible} energy carried away essentially by muons and neutrinos. This correction is of about $10$\% for a $10^{19}$ eV proton induced shower (simulated with QGSJet01), and it decreases as the shower energy increases \cite{invE}.

A fit of the correlation between the SD signal and the FD energy provides the energy calibration of the SD. An example of a calibration curve as obtained for 795 high-quality hybrid events is shown in Figure 1 (left panel) \cite{ClaudioICRC}.

\section{Energy spectrum}

\begin{figure}[htbp]
  \begin{center}
   \includegraphics[width=0.99\textwidth]{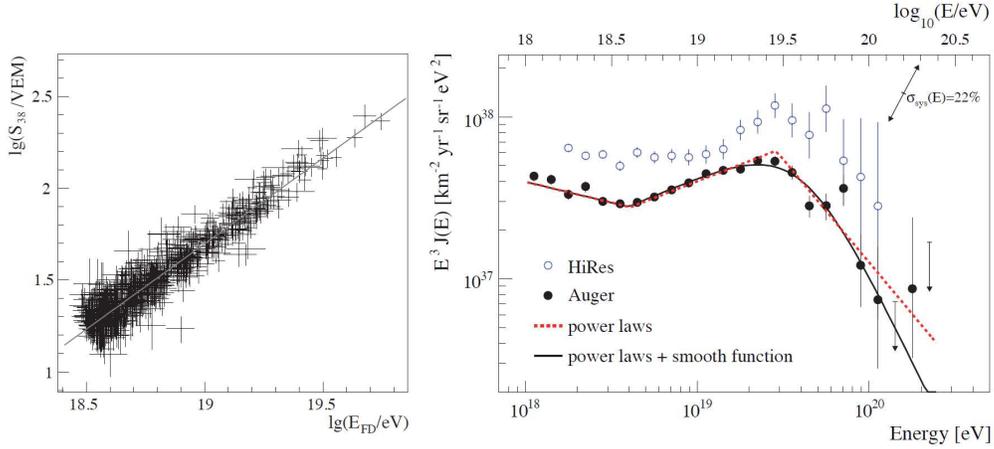}
   \caption{\emph{Left:} Example of a calibration curve obtained from a sample of 795 high-quality hybrid events, used to calibrate the SD signal ($S_{38^\circ}$) to the FD energy \cite{ClaudioICRC}. \emph{Right:} Combined hybrid and SD energy spectra from the Auger Observatory \cite{AugerSpectrum1,AugerSpectrum2} compared to the stereo spectrum from HiRes \cite{HiResSpectrum}.}
   \label{fig:spectrum}
 \end{center}
\end{figure}

The Pierre Auger energy spectrum is presented in figure \ref{fig:spectrum} (right panel) \cite{AugerSpectrum1,AugerSpectrum2}. The higher energy points are obtained with the SD while the lower energy points, near $10^{18}$ eV, are from the hybrid events, since in this region the SD is not fully efficient. The energy spectrum reflects the high statistics, enough to clearly identify the spectrum features, namely the so-called ankle region and a flux suppression at the highest energies. The fit to a broken power law indicates that the ankle is located at $\log(E/eV) =18.61\pm0.01$.

The significance of the flux suppression at the highest energies is greater than $20\ \sigma$. Although this suppression is expected due to UHECRs interactions with the Cosmic Microwave Background (CMB), through photo-pion production (the so-called GZK effect \cite{GZK1,GZK2}) or by photo-disintegration (for nuclei), one should keep in mind that it also may be caused by the exhausting of the acceleration mechanisms at the sources.
The HiRes result \cite{HiResSpectrum} is also displayed in Figure \ref{fig:spectrum} for comparison. Although the statistical uncertainty is larger than for the Auger Observatory, both energy spectra display essentially the same features and are in agreement within the systematic uncertainties ($22$\%).

\section{Mass Composition}

The knowledge of the UHECR composition is another key aspect not only to understand the acceleration mechanisms at the sources but also to be able to characterize the first interaction.

The depth of the shower maximum, $X_{max}$, is sensitive to the type of primary particle. Proton initiated showers, having a lower cross-section (deeper first interaction point) and spending their energy at a slower rate than iron showers, have in average a higher $X_{max}$, and larger fluctuations from event to event. As mentioned before, the longitudinal development of the shower can be followed with the Fluorescence Detector. The maximum depth of the shower can then be extracted by fitting a Gaisser-Hillas function to the profile. Although conceptually simple, the procedure to obtain the average $X_{max}$ and its fluctuations ($RMS$) is much more complex \cite{Unger}. Firstly, one has to apply quality cuts to the samples, ensuring a good reconstruction of the profile and estimating additional contributions coming from the Cherenkov beamed light. Afterwards, one has to ensure that the obtained $X_{max}$ distribution did not get biased by all the previous cuts, and due to the detector aperture. For this purpose, anti-bias cuts have to be applied. In the Auger Observatory these cuts are derived from data by selecting shower geometries that do not create a bias on the average $X_{max}$.

\begin{figure}[htbp]
  \begin{center}
   \includegraphics[width=0.99\textwidth]{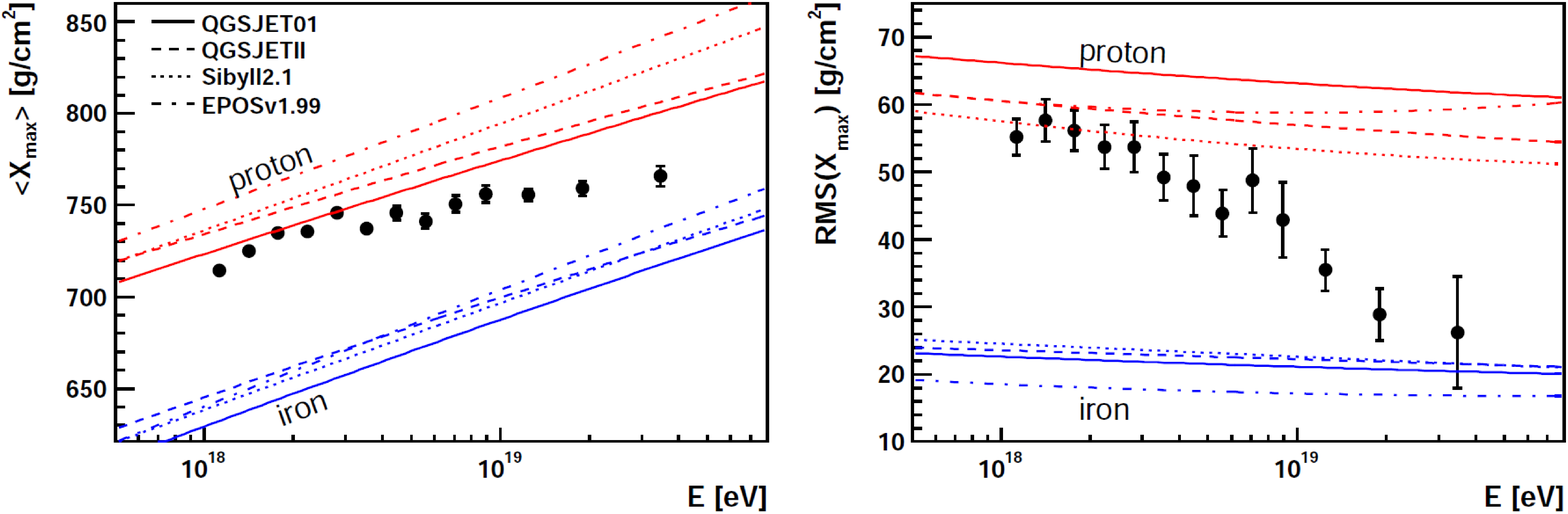}
   \caption{Measurements of $\left< X_{max} \right>$ (\emph{left}) and its $RMS$ (\emph{right}) as a function of energy \cite{AugerXmax}. The lines represent the predictions obtained from shower simulations based on different hadronic interaction models.}
   \label{fig:Xmax}
 \end{center}
\end{figure}

The obtained results on the $\left< X_{max} \right>$ and its fluctuations ($RMS$) are shown in Figure \ref{fig:Xmax} \cite{AugerXmax}. The predictions of these quantities as obtained from simulations of proton and iron showers and adopting different hadronic interaction models, are also shown in Figure \ref{fig:Xmax}. The data may suggest a transition from a lighter composition to a heavier one, although the $RMS$ seems to have a faster transition. At this point, it is important to note taht the measured $\left< X_{max} \right>$ and its fluctuations reflect both the primary composition of the observed showers and their physical cascade processes. The interpretation of these parameters thus depends also on assumptions made about those pocesses, in particular on the assumed high energy hadronic interaction model.

\section{High energy hadronic interaction models}

The hadronic interactions at high energy are a very interesting subject by itself. For the shower development the most important region is the forward one. Here the strong coupling constant is very high, which means that there is no asymptotic freedom of the partons and therefore perturbative-QCD cannot be applied. This is the soft regime and there is no established theory that can fully describe interactions in this region. Thus, the hadronic interactions at high energies are described through phenomenological models (that make use of the Gribov-Regge theory) that are fitted to the available accelerator data and extrapolated several orders of magnitude to the UHECRs energies.

\begin{figure}[htbp]
  \begin{center}
   \includegraphics[width=0.7\textwidth]{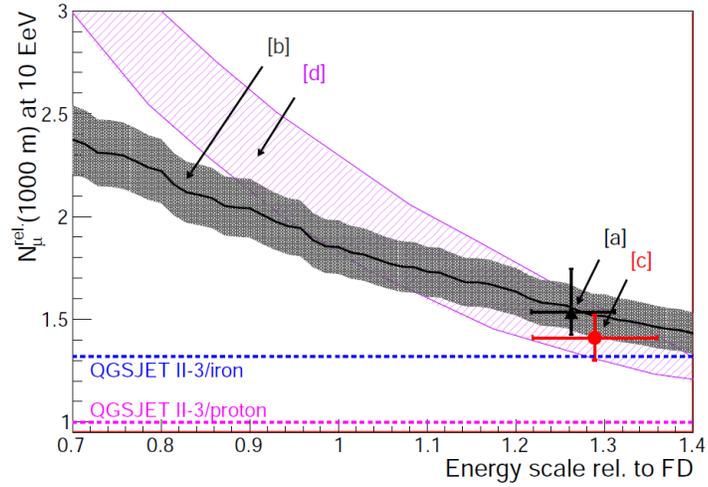}
   \caption{Number of muons, at $1000$ m from the shower core, relative to QGSJET-II/proton vs. the energy scale from [a] the universality method (triangle); [b] the jump method (filled area); [c] the smoothing method (circle); [d] the golden hybrid analysis (dashed area). The data have been selected for $log(E/eV) = 19.0 \pm 0.02$ and $\theta \le 50^\circ$. According to the tested model, Iron primaries give a number of muons $1.32$ times higher than that from protons (horizontal lines in the figure) \cite{AugerMuons}.}
   \label{fig:muons}
 \end{center}
\end{figure}

The number of muons provides a tool to test hadronic interaction models. Muons in EAS are mainly the sub-product of charged pions, and once they are produced, they have a large probability to reach the ground without decaying or interacting. Hence, they are intimately connected to the hadronic shower development.

In the Pierre Auger Observatory the number of muons can be obtained directly for inclined showers\footnote{this subject was addressed in the talk by R. A. V\'azquez given in this conference} or in an indirect way for vertical showers $(\theta < 50^\circ)$. In Figure \ref{fig:muons} the number of muons estimated by the different methods with respect to QGSJET-II/proton as function of the energy scale of the SD relative to the FD is shown \cite{AugerMuons}.

Two of the methods ([b] and [c]) analyse the signals given by the PMTs (FADC traces) to either identify the number of muons, and this is done counting the peaks (corresponding to muons) - the jump method, or by estimating the electromagnetic signal in the tank, by taking into account that it should be a smooth signal - this method is called the smoothing method.
Another method ([a]) makes use of hybrid events and the universality of the electromagnetic signal at ground. Finally the method called golden hybrid ([d]) uses events that can be reconstructed independently by SD and FD. Then, simulated events are fitted to the recorded longitudinal profile and the corresponding signal at ground level found in simulation is compared with the one in data. 

All the four methods predict less muons resulting from the application of the hadronic interaction models than measured. A higher energy scale of the SD is also favoured. This last result is compatible with the systematic uncertainty assigned to the energy as given by the FD.

\section{Final Remarks}

The Pierre Auger Observatory baseline configuration is complete and is now running smoothly.
It acquired enough statistics to obtain the energy spectrum where both the ankle structure and a GZK-\emph{like} suppression are clearly observed.
The $\left< X_{max} \right>$ and its $RMS$ may suggest a transition on the primary mass composition from light to heavy. However, the interpretation of these observables depends also on the high energy hadronic interaction models. These have been tested in Auger through the measurement of the number of muons. The obtained results show that all the models predict less muons at the ground than measured, even for iron induced shower.

The Pierre Auger Observatory offers an unique window to explore particle physics at centre-of-mass energies about one order of magnitude higher than those reachable with the present technology. It is also exploring a kinematical region usually inaccessible to man-made accelerators, the forward region. Furthermore, the Large Hadron Collider (LHC) is providing data that can constrain the hadronic interaction models ($\sim 7$ TeV).

Finally, the Pierre Auger Observatory is accumulating statistics which will allow, in a near future, more sophisticated analysis, that could solve the UHECRs puzzle.

\section*{Acknowledgments} 
I want to thank my colleagues from the Pierre Auger Collaboration for reading the proceeding, and to acknowledge the financial support given by FCT, Funda\c{c}\~{a}o para a Ci\^{e}ncia e Tecnologia (SFRH/BPD/73270/2010).

\end{document}